\documentclass[aps,twocolumn]{revtex4-1}
\usepackage{newlfont}
\usepackage{color}           
\usepackage{soul}            
\usepackage{amssymb}
\usepackage{amsfonts}
\usepackage{amsmath}
\usepackage{wasysym}
\usepackage{graphicx}
\usepackage{epsfig}
\usepackage{amsthm}
\usepackage{bm}
\usepackage[colorlinks=true, citecolor=blue, urlcolor=blue ]{hyperref}

\begin{document} 
 
\title{Limit on Time-Energy Uncertainty with Multipartite Entanglement}
\author{Manabendra Nath Bera, R. Prabhu, Arun Kumar Pati, Aditi Sen(De), Ujjwal Sen}
\affiliation{Harish-Chandra Research Institute, Chhatnag Road, Jhunsi, Allahabad 211 019, India}

\begin{abstract}
 We establish a relation between the geometric time-energy uncertainty and multipartite entanglement. In particular, we show that the time-energy uncertainty relation is bounded below by the geometric measure of multipartite entanglement for an arbitrary quantum evolution of any multipartite system. 
The product of the time-averaged speed of the quantum evolution and the time interval of the evolution is bounded below by the multipartite entanglement of the target state. This relation holds for pure as well as for mixed states. We provide examples of physical systems for  which the bound reaches close to saturation.

\end{abstract}

\maketitle

\section{Introduction}  
In the beginning of the last century, the geometry of space-time has played an important role in the reformulation of classical  mechanics. 
Similarly, it is hoped that a geometric formulation of quantum theory, and an understanding of the geometry of quantum state space 
can provide new insights into the nature of quantum world and, in particular, into the field of quantum information science.
In quantum theory, the geometric properties of the quantum state space are characterized by the Riemannian metrics 
 defined on the corresponding projective Hilbert space \cite{Provost80,Anandan90,Anandan90a,Anandan91,Pati91,Braunstein94}. 
The discovery of geometric phase, by Pancharatnam \cite{Pancha56} and then by Berry \cite{Berry84},
provided a new fillip to the importance of geometric properties in the quantum domain. The connection of this phase to the 
Riemannian metrics lead to a new understandings in terms of the geometrical ideas on the projective Hilbert space  \cite{Simon83}.
One of the significant outcomes of this geometric approach is the geometric quantum uncertainty relation (GQUR) 
\cite{Anandan90,Anandan91,Uhlman92} which constraints the motion of a system in its projective Hilbert space.
Importantly, it was found  
that for any such evolution, there exists a lower bound on the product of the time separation between initial and final state and its 
time-averaged energy fluctuation \cite{Anandan91}. 
%





Entanglement has played a pivotal role in the 
recent developments of quantum information \cite{Horodecki09}.
On the practical side, the major challenge is to create, store, and process multiparty entanglement in a controlled manner. 
This has created an immense interest to understand the properties of entangled quantum states and the dynamics of entanglement in a
multiparticle scenario \cite{Sen007}.
An unentangled multiparty quantum system, if allowed to interact,  may evolve to an entangled state, 
depending on the initial state and the driving Hamiltonian.
Hence it is important to  identify the class of initial states and the class of  non-local
Hamiltonians that can create a desired entangled quantum state. 
The entangling capacity of Hamiltonians for two qubits and the relation of the ``speed'' of the quantum evolution with the 
entanglements of the initial and final states were addressed in Ref. \cite{Dur01}.

In this article, we pose the question: does multipartite entanglement have any
connection to the geometry of quantum evolution? 
Or, more generally, can the geometric quantum uncertainty relation be
connected, quantitatively, with the multipartite entanglement present in
the system? This question is important not only due to its fundamental nature, but
also because of its practical relevance in quantum information.
We establish a relationship between the multipartite entanglement in a many-body quantum
system and the total distance traveled
 by the state (pure or mixed) during its evolution. 
This leads us naturally to consider the 
geometric time-energy uncertainty relation for multiparty quantum systems, and
to provide a connection of the same to multipartite entanglement. 
In particular, we find that the minimum time taken by an initial quantum state (pure or mixed) to reach a target entangled state
is bounded below by the geometric measure of multipartite entanglement \cite{GMrefs,GGM} (cf. \cite{sobmulti} ) upto a factor depending 
upon the corresponding path-integrated energy fluctuation.
In the case of mixed states, we show that the relation is generic, in that its form is independent of the specific metric used. 
We exemplify this by using various metrics such as,
the Fubini-Study \cite{Anandan90, Pati91}, the Hilbert-Schmidt \cite{Anandan91}, and the Bures metrics \cite{Uhlman92,Braunstein94}. 
We then investigate the relation of genuine multipartite entanglement with the GQUR by considering a physically realizable quantum 
many-body model, viz., the Heisenberg spin chain.


The structure of the article is as follows. In Sec. II, we introduce the geometric quantum uncertainty relation (GQUR) in connection to quantum evolution for pure states.  We derive the relation between the GQUR and the multipartite entanglement measures for the corresponding pure states, and illustrate the relation with examples, in Sec. III.
In Sec. IV, we show that a similar connection  can be obtained for mixed quantum states using different metrics. Finally, we conclude in Sec. V.  

\section{Geometric quantum uncertainty relation}
The Hilbert space formalism of quantum mechanics possesses interesting geometric properties, and their inherent features are used to build the geometric quantum uncertainty relation \cite{Provost80,Anandan90,Anandan90a,Anandan91,Braunstein94,Zyczkowski09}.
Let $\{| \Psi \rangle \}$ be a set of vectors in a multiparticle Hilbert space ${\cal H} = \otimes_{i=1}^N {\cal H}_i $. Since two unit vectors differing only in phase, represent the same physically (i.e., quantum mechanically) equivalent states, one can construct a ray space from these equivalent classes of states.
The set of rays of ${\cal H}$, via a projection map, forms the projective Hilbert space  
${\cal P}({\cal H})$. If dim${\cal H}_i= d$ $\forall i$, then ${\cal H} \cong \mathbb{C}^{d^N}$. Correspondingly, for the projective Hilbert space  $\cal P = {\cal P(\cal H)} \cong  (\mathbb {C}^{d^N} - \{0\})/U(1)$, which is a complex manifold of dimension $(d^N -1)$. This can also be considered as a real manifold 
of dimension $2(d^N -1)$. Any quantum state at a given instant of time can be represented as a point in ${\cal P}$ via
the projection map $\Pi: |\Psi\rangle \rightarrow |\Psi\rangle \langle \Psi|$. The evolution of the state vector represents a curve
$C: t \rightarrow |\Psi(t)\rangle$ in ${\cal H}$ whose projection $\Pi(C)$ lies in ${\cal P}$ 
\cite{Provost80,Simon83,Anandan90,Anandan90a,Anandan91,Pati91,Braunstein94,Zyczkowski09}. 

For a given curve, one can ask how much distance the system has traveled  during its time evolution. As the state evolves under the 
Schr\"odinger evolution, it traces a curve in the Hilbert space. Different Hamiltonians may give different curves, but all of them may 
be projected to a single curve in ${\cal P}$.
For pure states, the distance, $S$, between any two points in the projective Hilbert space $\mathcal{P}$, corresponding to the
quantum states $| \Psi (\overline{\lambda}) \rangle$ and $| \Psi (\overline{\lambda} ^{\prime}) \rangle$ can be defined with the Fubini-Study distance \cite{Anandan90}, as
\begin{equation}
 S^2=4 \left(1 - | \langle \Psi (\overline{\lambda}^{\prime}) |\Psi (\overline{\lambda}) \rangle |^2 \right),
 \label{eqn:fsd}
\end{equation}
where $\overline{\lambda}$ and $\overline{\lambda}^{\prime}$ are parameters on which these states depend. The distance is in fact between rays corresponding to the vectors of $\cal H$, and is hence defined in $\cal P$.
If two vectors differ from each other infinitesimally then we have the  infinitesimal Fubini-Study metric as given by
\begin{equation}
\begin{split}
 dS^2 & = 4 \left(1-| \langle \Psi (\overline{\lambda} + d \overline{\lambda}) | \Psi (\overline{\lambda}) \rangle|^2 \right) \\
      & = 4 \left( \langle \partial_i \Psi | \partial_j \Psi \rangle- \langle \partial_i \Psi |  \Psi \rangle  \langle  \Psi | \partial_j \Psi \rangle \right) 
d \lambda^i d\lambda^j, 
\end{split}
\label{eqn:fsmetric}
\end{equation}
where we have retained terms upto second order only.
Another distance, denoted as $\mathcal{S}$ can be defined via the Bargmann angle \cite{Anandan90}
\begin{equation}
 |\langle \Psi_1|\Psi_2\rangle |^2=\cos^2\left(\frac{\mathcal{S}}{2}  \right).
 \label{eqn:spherDis}
\end{equation}
For infinitesimally close vectors, the Fubini-Study distance, and the distance obtained via the Bargmann angle are the same and is given by Eq. (\ref{eqn:fsmetric}). In addition, one can also define the ``minimum normed distance'' between the states 
$| \psi_1 \rangle$ and $|\psi_2 \rangle$ 
as \cite{Provost80,Pati91}
\begin{equation}
 S_N^2=2 \left(1 - |\langle \Psi_1| \Psi_2 \rangle |  \right).
 \label{eqn:mnd}
\end{equation}
However, when the states differ infinitesimally, we get $4dS_N^2=dS^2$.

Let us now consider the parametrization of a curve in $\mathcal{H}$ using the time $t$. 
During the unitary time evolution of the quantum state by the Hamiltonian $H$,
the path taken by the state should be guided by the property of the quantum state and the characteristics of the Hamiltonian.
The important result obtained by  Aharonov 
and Anandan is that \textit{the energy fluctuation of the state
drives the quantum evolution} \cite{Anandan90}. In the projective 
Hilbert space $\mathcal{P}$, an isolated system can move if and only if it is not in a stationary
state, i.e., it has a non-zero energy fluctuation, which for the state $| \psi(t) \rangle$ is denoted as $\varDelta H(t)$.

Suppose that a quantum state $ | \Psi(t) \rangle$ is transported to a state $ |\Psi(t+dt) \rangle$ after an 
infinitesimal time interval $dt$, following the Schr\"odinger evolution governed by the Hamiltonian $H(t)$.
By using the Fubini-Study metric in Eq. (\ref{eqn:fsmetric}), the infinitesimal distance that the system traverses during this 
evolution is given by \cite{Anandan90}
\begin{equation}
 dS=\frac{2}{\hbar} \varDelta H(t) dt,
\end{equation}
where $\varDelta H(t)$ is the energy fluctuation, defined by
 \begin{equation}
  \varDelta H(t)^2 = \langle \Psi(t) | H(t)^2 | \Psi(t) \rangle - \langle \Psi(t) |
H(t) | \Psi(t) \rangle^2.
 \end{equation}
The above relation implies that a higher energy fluctuation leads to a higher speed in the evolution.
If the initial state $|\Psi(0) \rangle$ is transported to the state $| \Psi(\tau) \rangle$ after a time $\tau$,
it follows a path whose total distance is
\begin{equation}
  S= \frac{2}{\hbar} \int_0^{\tau}   \varDelta H(t) dt.
 \end{equation}
%
%
For a time independent Hamiltonian, the energy uncertainty, $\varDelta H$, is a constant throughout the evolution. In this 
case, the total distance that the system traverses during its evolution, from $| \Psi(0)\rangle$ to $| \Psi(\tau)\rangle$, is given by
\begin{equation}
S = \frac{2}{\hbar} \tau \ \varDelta H .
 \label{eqn:taa}
\end{equation}
Note here that there are infinitely many Hamiltonians can be used to transport the same initial state $| \Psi(0)\rangle$ to the same final state $| \Psi(\tau)\rangle$. The distance traversed for any such path must be lower than $S_0$, where $S_0$ is
the shortest geodesic connecting $| \Psi(0)\rangle$ and $| \Psi(\tau)\rangle$ in $\mathcal{P}$, according to the same metric that is used to calculate the actual distance traveled in the quantum evolution.
In other words, we have
\begin{equation}
 \tau \varDelta H \geqslant \hbar \cos^{-1} (|\langle \Psi(0)|\Psi(\tau)\rangle |).
 \label{eqn:gur}
\end{equation}
where we have used the metric generated via the Bargmann angle. One can also use the Fubini-Study distance of Eq. (\ref{eqn:fsd}) or the minimum normed distance of Eq. (\ref{eqn:mnd}) to generate the metric, in which case weaker relations are obtained. Such relations are usually referred to as the geometric quantum uncertainty relation (GQUR). The equality in Eq. (\ref{eqn:gur}) holds only for the quantum trajectories that coincide with the geodesics in $\mathcal{P}$.  Quantum states which satisfy the equality are known as intelligent states \cite{mann}. In case, the Hamiltonian is time-dependent, the GQUR is
\begin{equation}
 \int_0^\tau \varDelta H (t) dt \geqslant \hbar \cos^{-1} (|\langle \Psi(0)|\Psi(\tau)\rangle |).
 \label{eqn:gqur1}
\end{equation}

\section{GQUR and multipartite quantum entanglement}
Entanglement measures of multiparty quantum states can be
defined by considering the geometric distance as measured by a given metric 
from a specified set of non-entangled (separable) quantum states, in the Hilbert space. Depending on the set of non-entangled states, from which the distance is 
measured, the distance quantifies different types of entanglement present in the system.
For an N-party pure quantum state $| \Psi_N \rangle$, such a  measure of multipartite entanglement can be defined as \cite{GMrefs,GGM}
\begin{equation}
 \mathcal{E_{\mathcal{C}}}(| \Psi_N \rangle)=\min_{\{| \Phi_m\rangle \in S_{\mathcal{C}} \}} \left(1- |\langle \Phi_m|\Psi_N\rangle|^2\right),
 \label{eqn:pGM}
\end{equation} 
where the minimization is carried over a certain set of separable states $S_{\mathcal{C}}$. There is a hierarchy of such geometric 
measures, depending on the set of states $\{| \Phi_m \rangle \in S_{\mathcal{C}} \}$ chosen for the minimization. Among the set of such  multipartite entanglement measures, $\mathcal{E_{\mathcal{C}}}$, two are more prominent, in that they are defined with respect to two extremal choices of the set $S_{\mathcal{C}}$. More precisely, choosing $S_{\mathcal{C}}$ to be the set, $S_0$, of all fully separable states, i.e., states of the form $| \Phi^1 \rangle \otimes| \Phi^2 \rangle \otimes  \hdots \otimes | \Phi^N \rangle$ in the N-party tensor product Hilbert space of the N-party system, we obtain the ``geometric measure'' (GM) of multiparty entanglement \cite{GMrefs}. We denote it by $\mathcal{E}_0$.
On the other hand, if the set $S_{\mathcal{C}}$ is formed by the states which are not genuinely multipartite entangled, 
the ``generalized geometric measure'' (GGM), 
is obtained \cite{GGM}. An N-party pure quantum state is said to be genuinely multiparty entangled if it is entangled across every partition of the N parties into two disjoint sets. We denote this set of separable set as $S_G$ and the GGM as $\mathcal{E}_G$.
In general, if the minimization is carried out over the set of $(k-1)$-separable states, the measure will be called the geometric measure $\mathcal{E}_{N-k+1}$. Note that the GGM, $\mathcal{E}_G$, gives the lowest value among all the geometric multipartite entanglement measures,  for a given multipartite quantum state.
We can define another geometric multipartite  entanglement  measure, $\mathcal{G}\left(\mathcal{E}_{\mathcal{C}} \right)$, which is a monotonically increasing function of $\mathcal{E}_{\mathcal{C}}$, as
 \begin{equation}
  \mathcal{G}\left(\mathcal{E}_{\mathcal{C}} \right)= \cos^{-1} \sqrt{  1-\mathcal{E}_{\mathcal{C}} }.
  \label{eqn:ge}
 \end{equation}
Note that $\mathcal{G}\left(\mathcal{E}_{\mathcal{C}} \right)$ also satisfies all the properties of entanglement measures which $\mathcal{E}_{\mathcal{C}}$ satisfies.

Let us now build the connection between the geometric quantum uncertainty relation and the geometric measure of entanglement, 
$\mathcal{E}_{\mathcal{C}}$. 

\noindent \textbf{Theorem:} \emph{For an arbitrary quantum evolution, the time interval multiplied by the time-averaged energy fluctuation is bounded below by the geometric measure of multipartite entanglement, provided the initial state is unentangled.}

\noindent \texttt{Proof:} By definition, we have 
\begin{equation}
 |\langle \Psi(0)|\Psi(\tau)\rangle |^2 \leqslant 1-\mathcal{E}_{\mathcal{C}} \left(|\Psi(\tau)\rangle \right)
\end{equation}
With the definition in Eq. (\ref{eqn:ge}) and remembering that the distance function is always positive, the above relation becomes 
\begin{equation}
 \cos^{-1} \left( |\langle \Psi(0)|\Psi(\tau)\rangle | \right) \geqslant \mathcal{G}\left(\mathcal{E}_{\mathcal{C}} \right).
\label{eqn:cos}
 \end{equation}
Now, using the geometric quantum uncertainty relation in Eq. (\ref{eqn:gqur1}) and the above equation, we obtain
\begin{equation}
 \overline{\varDelta H} \tau  \geqslant \hbar \ \mathcal{G}\left(\mathcal{E}_{\mathcal{C}} \right),
\label{eqn:gqur}
\end{equation}
where $\overline{\varDelta H}$ is the time-averaged energy fluctuation, defined as
\begin{equation}
 \overline{\varDelta H}=\frac{1}{\tau}\int_0^{\tau} \varDelta H(t) dt,
\end{equation}
which can also be interpreted as the time-averaged speed of the evolution.
The equality in Eq. (\ref{eqn:gqur}) holds only when the Hamiltonian which drives the state follows the geodesic, and
the  $|\Psi(0)\rangle$ coincides with the state $| \Phi_{\min} \rangle$ for which the minimization is 
achieved in $\mathcal{E}_{\mathcal{C}}$. \hfill $\blacksquare$

Note here that the ``unentangled'' initial state in the statement of the theorem is from the same set of separable states, $S_{\mathcal{C}}$, which is used in the definition of the $\mathcal{E}_{\mathcal{C}}$.

The theorem implies that for a given energy uncertainty $\varDelta H$, the minimum time $\tau$ required to connect the initial and final states depends on the entanglement present in the final state. It requires more time, $\tau$ when the final state has more 
multipartite geometric entanglement. 
This result is potentially important in foundational as well as practical aspects of quantum information and beyond. 


Let us now illustrate the effectiveness of the above bound by considering some specific Hamiltonians. 
Consider the $N$-qubit product state in the standard product form given by
\begin{equation}
|\Psi\rangle=|\varphi\rangle^{\otimes N},
\label{eq:state}
\end{equation}
where $|\varphi\rangle=\cos\theta|0\rangle + \mbox{exp}(-i\phi) \sin\theta |1\rangle$ with $\theta\in[0,\pi]$ and $\phi\in[0,2\pi)$, and where $|0\rangle$ and $|1\rangle$ are eigen vectors of the $\sigma_z$ Pauli matrix with the eigen values +1 and -1, respectively. 
Suppose that the evolution of the state is guided by the Ising Hamiltonian
\begin{equation}
H_I = \frac{J}{4}\left(\sum_{i=1}^{N-1} (I-\sigma^z_i)(I-\sigma^z_{i+1})\right).
\label{eq:Ising}
\end{equation}
We are therefore considering a liner arrangement of $N$ quantum spin-1/2 particles.
$\sigma_i^a \ (a=x,y,z)$ are the Pauli spin matrices at the site $i$. $J$ is a system parameter having the dimension of energy. This Hamiltonian has been used to obtain the ``cluster state'' which is an important substrate for quantum computation \cite{Briegel09}.

Starting from the initial state $|\Psi\rangle$ with $\phi=0$, we now check whether the lower bound given in Eq. (\ref{eqn:gqur}) 
for this Hamiltonian is tight or not. To do this, we generate all possible initial states by varying the state parameter $\theta$, and evolve the system for each such initial state, by using the Ising Hamiltonian.

\begin{figure}[h]%
\includegraphics[width=7cm]{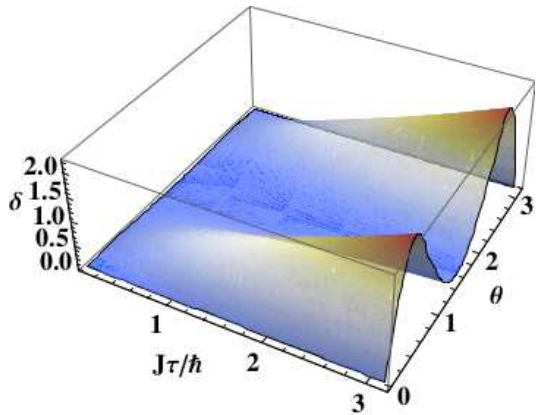}%
\caption{(Color online) Plot of $\delta=\frac{\tau \Delta H}{\hbar} - \cal{G}(\cal{E}_G)$ with respect to $\frac{J \tau}{\hbar}$ and initial state parameter $\theta$. 
The product state, given by Eq. (\ref{eq:state}), evolves according to the  cluster Hamiltonian, given in Eq. (\ref{eq:Ising}).
The bound is tight for small evolution time. While $\delta$ and $\frac{J \tau}{\hbar}$ are dimensionless parameters, $\theta$ is measured in radians.}
\label{fig:figure1}%
\end{figure}

For ease of demonstration, we set
\begin{equation}
\delta=\frac{\tau \Delta H}{\hbar} - \cal{G}(\cal{E}_G). 
\end{equation}
So the entanglement bound of the time-energy geometric uncertainty is saturated if $\delta=0$.
In Fig. \ref{fig:figure1}, we plot $\delta$, in which the evolution is governed by the Hamiltonian given in Eq. (\ref{eq:Ising}), consisting of three spins, with respect to the evolution time $\tau$ and the initial state parameter $\theta$. 

\begin{figure*}[htbp]
\centering
\includegraphics[width=0.25\textwidth]{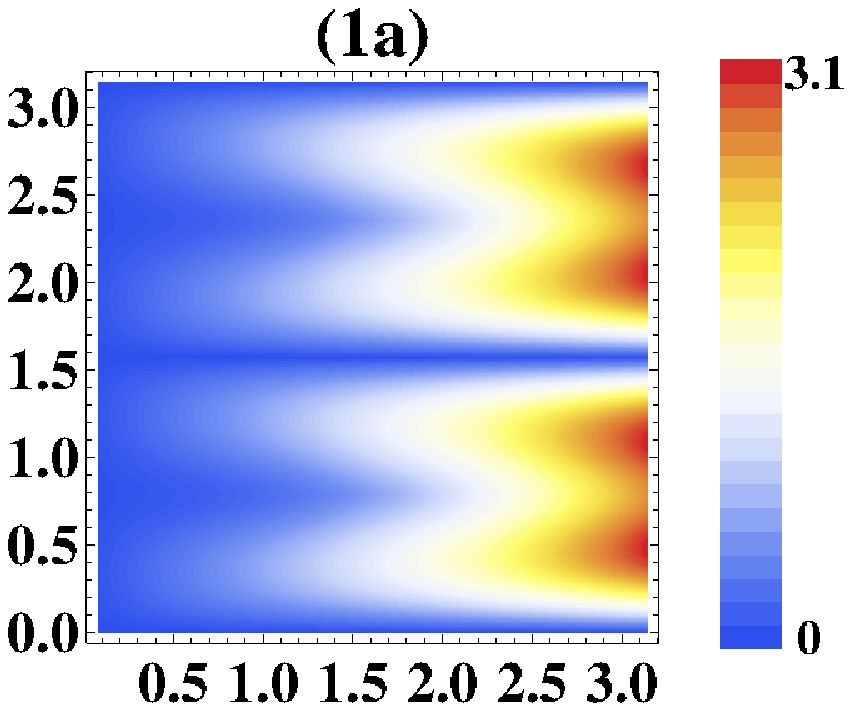}
\includegraphics[width=0.25\textwidth]{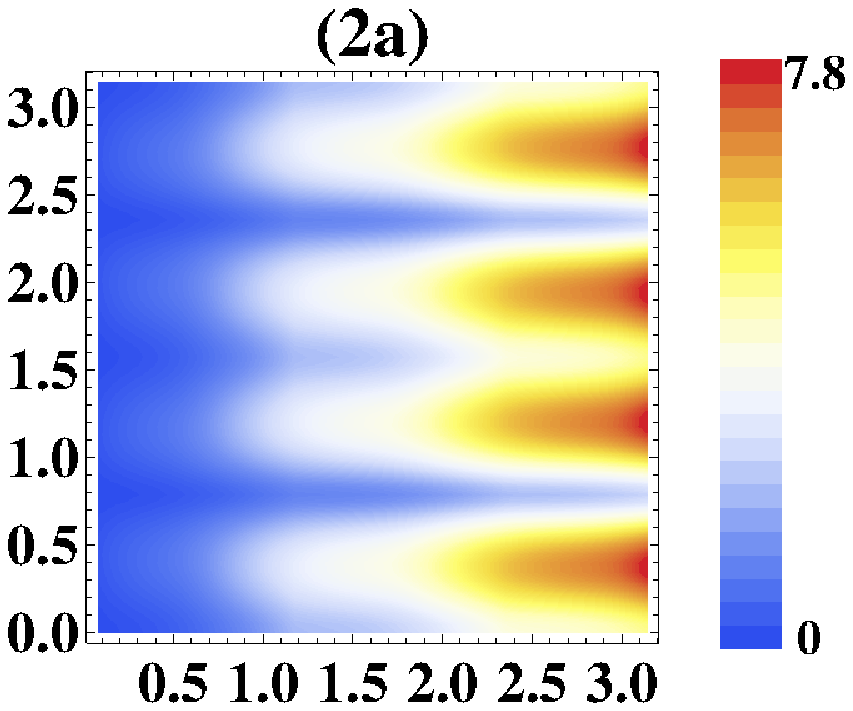}
\includegraphics[width=0.25\textwidth]{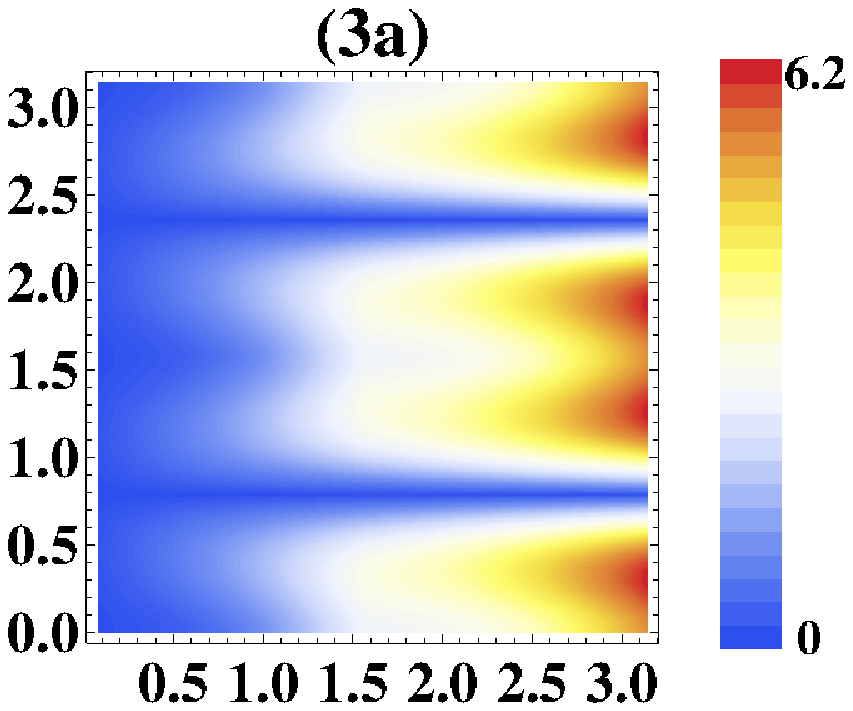}
\includegraphics[width=0.25\textwidth]{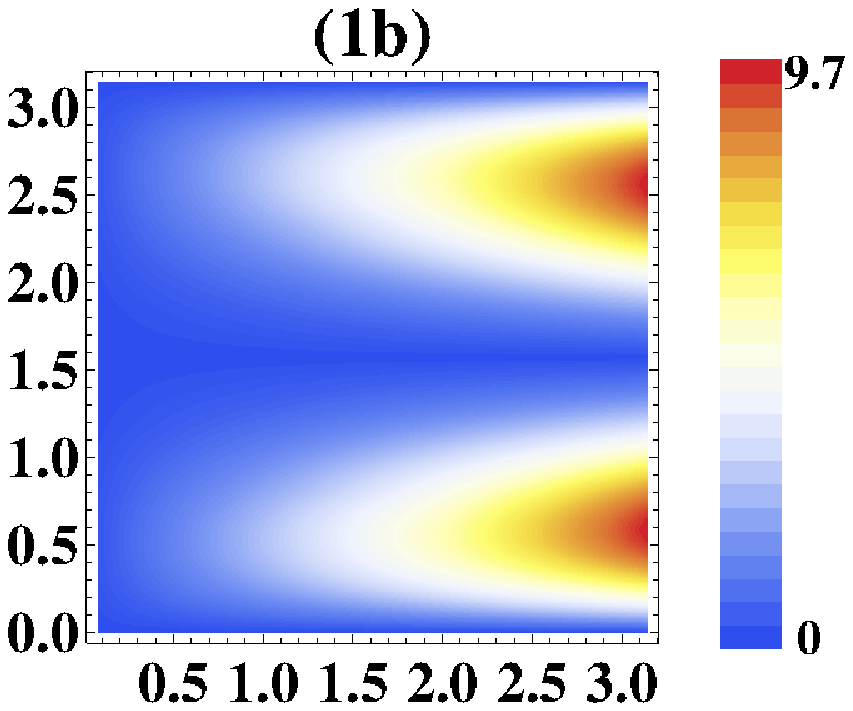}
\includegraphics[width=0.25\textwidth]{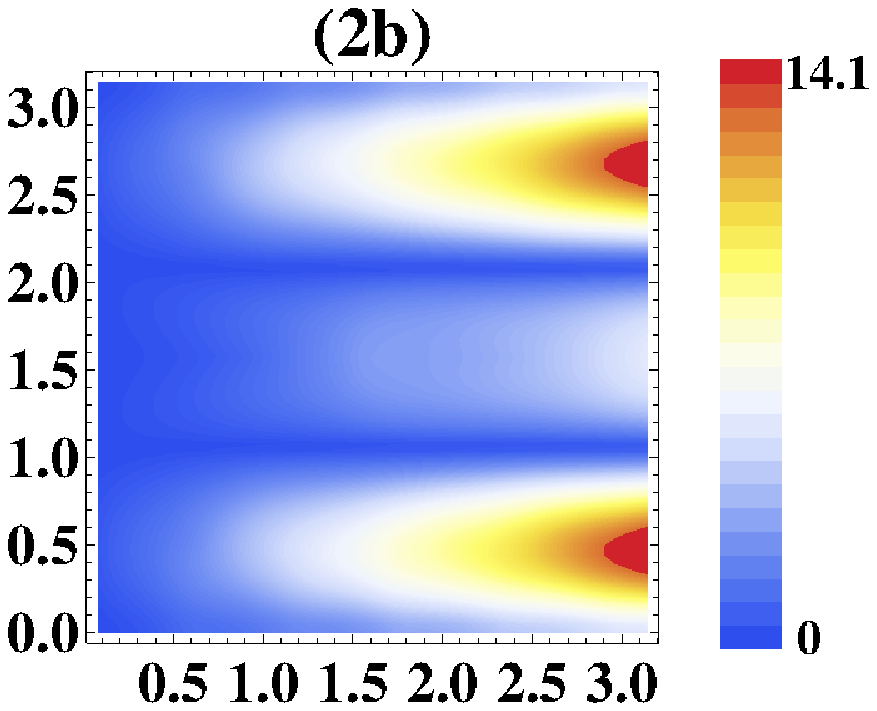}
\includegraphics[width=0.25\textwidth]{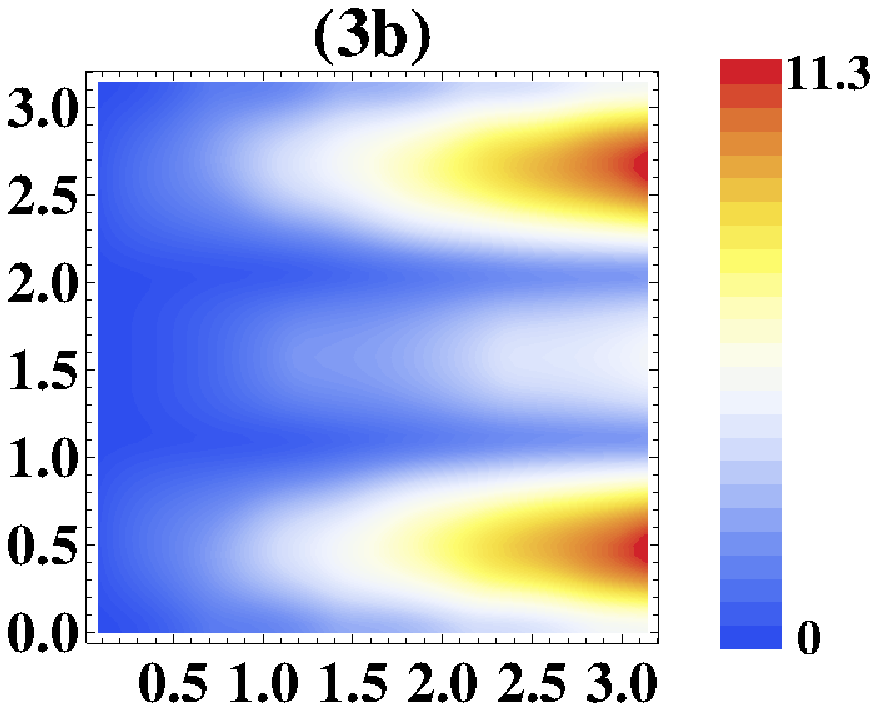}
\caption{
(Color online) Projection plots of $\delta$ with respect to the evolution time $\frac{J \tau}{\hbar}$ (horizontal axes) and initial state parameter 
$\theta$ (vertical axes). (a) and (b) respectively represent $\delta$ with no magnetic field (i.e., $h=0$) and with $h=1.5$. (1) 
is the plot of $\delta$ for  $\gamma=0\, \mbox{and}\, \mu=0.5$, (2) corresponds to the same for $\gamma=0.5\, \mbox{and}\, \mu=0$, and (3) represents  the same for $\gamma=0.5\, 
\mbox{and}\, \mu=0.5$. Clearly, in all the cases, the bound is tighter for low $\tau$ than for larger times. However, at large time, 
when the applied magnetic field is high, then the bound is more effective as compared to low magnetic fields. This feature remains the same for other choices of the parameters $\gamma,\, \mu,\, \mbox{and}\, h$.
The units are as in Fig. \ref{fig:figure1}.}
\label{fig:figure2}%
\end{figure*}

To investigate whether the saturation of the   lower bound is potentially generic, we evolve the state, given in Eq. (\ref{eq:state}), by considering the anisotropic XYZ (Heisenberg) Hamiltonian, given by
\begin{eqnarray}
H_H = J \sum_{i=1}^N \left[ (1+\gamma)\sigma^x_i \sigma^x_{i+1} + (1-\gamma)\sigma^y_i \sigma^y_{i+1} \right.  \nonumber \\
  \ \ \ \ \ \ \ \ \ \ \ \ \ \ \ \ \ \ \ \ \ \ \ \ \ \ \ \ \ \ \ \ \left. +\mu \sigma^z_i \sigma^z_{i+1}  + h \sigma^z_i \right]. 
\end{eqnarray}
Here,  $\gamma \in [0,1]$ is the anisotropy parameter which modulates the relative strengths of the $xx-$ and $yy-$interactions and $\mu$ determines the strength of the $zz-$interaction. And $h$ is the applied magnetic field. $\gamma, \ \mu$ and $h$ are the dimensionless system parameters. $J$ is a system parameter that has the dimension of energy. The Hamiltonian, $H_H$, again describes a system of $N$ quantum spin-1/2 particles arranged in a ring. The case for which $\gamma=1$ and $\mu=0$ corresponds to the transverse 
Ising Hamiltonian, while that for which $\gamma=0$ and $\mu=1$ corresponds to the isotropic Heisenberg Hamiltonian. By simulating the dynamics numerically, we observe that for any choice of  
$\gamma,\, \delta,\, \mbox{and}\, h$, the bound is tight for all low values of the time of evolution.  However,  there exist  
intermediate values of initial state parameter $\theta$, for which the bound, given in Eq. (\ref{eqn:gqur}), saturates also at large times of the evolution. Also, in the  ($\tau, \ \theta$)-space the saturation of the bound happens in a larger area for
 high magnetic fields, in comparison to low fields (see Fig. \ref{fig:figure2}). 

\section{Mixed states GQUR and multipartite entanglement}

Until now, we have dealt with the relation of the time-energy uncertainty with multipartite entanglement, for \emph{pure} quantum states. 
In this section, we show that it is also possible to obtain similar uncertainty relations for mixed multipartite quantum states, by again 
involving geometric multipartite entanglement measures.
Unlike for pure states, there are quantitatively distinct relations that are obtained for different metrics that can be utilized to metrize the space of density operators on the Hilbert space corresponding to the physical system under consideration. However, all metrics produce quantitatively similar relations. 


\subsection{GQUR and multipartite entanglement using Fubini-Study metric}
The distance between two arbitrary mixed quantum states can be quantified in several ways. One of the prominent ones is the
Hilbert-Schmidt distance, which is defined  for two arbitrary density matrices,
$\rho_1$ and $\rho_2$, as 
\begin{equation}
 S_{HS}\left[ \rho_1,\rho_2 \right]=\sqrt{ \mbox{Tr} \left( \rho_1 - \rho_2 \right)^2}.
\end{equation}
The above distance is Riemannian although it, in general,  is not  contractive under completely positive maps. If two density 
matrices, obtained via time evolution with the same Hamiltonian and from the same initial state,  differ infinitesimally in the time parameter $t$, the metric becomes
\begin{equation}
\begin{split}
 dS_{HS}^2 & =  \mbox{Tr} \left[ \rho (t+dt) - \rho(t) \right]^2  \\
           & = \mbox{Tr} \left( \dot{\rho}\right)^2 dt^2,
\end{split}
\end{equation}
where the $\dot{\rho}$ is the time-derivative of $\rho(t)$. The time evolution, governed by the Hamiltonian $H$, 
following the  Schr\"odinger-von Neumann equation, 
$i \hbar \dot{\rho}=\left[ H, \rho \right]$, leads to 
\begin{equation}
\label{HSmixed}
 \frac{dS_{HS}^2}{dt^2}=\frac{4}{\hbar^2} \mbox{Tr} \left[ (\rho^2 H^2) - (\rho H)^2  \right].
\end{equation}
The expression $2 \mbox{Tr} \left[ (\rho^2 H^2) - (\rho H)^2  \right]$ has been argued as an quantum mechanical fluctuation in energy. We denote it as $\varDelta H_Q(t)^2$ \cite{Anandan91}. A similar expression 
(up to a factor) also can be derived by using the Anandan distance, which is an extension of the 
Fubini-Study (FS) distance for mixed states \cite{Anandan91}. In fact, the metric, $dS_{FS}$, in this case, is related to $dS_{HS}$ as 
\begin{equation}
\label{FSvsHS}
\begin{split}
 dS_{FS}^2 & = 4 \left( 1-\frac{\mbox{Tr} \left[\rho(t+dt)\rho(t) \right]}{ \mbox{Tr} \left[\rho(t)^2 \right]}  \right) \\
           & = 2 \ dS_{HS}^2,
\end{split}
\end{equation}
where we assume that the Hilbert-Schmidt distance is normalized by the trace of the square of the density matrix at time $t$.  
Note here that the latter quantity (i.e., the normalization) is time-independent. For pure states, the above expression reduces to Eq. (\ref{eqn:fsmetric}). So, by using Eqs. (\ref{HSmixed}) and (\ref{FSvsHS}), we obtain a relation between the speed of the quantum evolution of the mixed quantum state with the quantum fluctuation in energy, as 
\begin{equation}
 \frac{dS_{FS}}{dt}=\frac{2}{\hbar} \varDelta H_Q(t).
\end{equation}
Therefore, if quantum fluctuation in energy is large, the speed of quantum evolution will be fast.
Following the arguments given in the case of pure quantum states, the geodesic distance $S_0$ measured between two mixed states, 
$\rho_1$ and $\rho_2$, with the help of the Fubini-Study metric, can be given by 
\begin{equation}
\frac{\mbox{Tr} \left[\rho_1 \rho_2 \right]}{\mbox{Tr}\left[\rho_1^2 \right]} = \mbox{cos}^2 \left( \frac{S_0}{2} \right).
\end{equation}
Let us now consider that 
an initial state $\rho(0)$ evolves to a final state $\rho(\tau)$ in the time interval $\tau$. In that case,  
the geometric uncertainty relation is given by
\begin{equation}
 \frac{1}{\hbar} \int_0^{\tau} \varDelta H_Q (t) dt \geqslant S_0 = 2 \mbox{cos}^{-1} \sqrt{\frac{\mbox{Tr} \left[\rho(0) \rho(\tau) \right]}{\mbox{Tr}\left[\rho(0)^2 \right]}}
\end{equation}
For a time-independent Hamiltonian $H$, the above inequality reduces to 
\begin{equation}
\tau \ \varDelta H_Q  \geqslant \hbar \ S_0.
\label{eqn:gqurmixfs}
\end{equation}

To relate the uncertainty relation 
to multipartite entanglement measures,
let us introduce a distance-based measure of quantum entanglement, based on the Fubini-Study metric.
For an arbitrary N-party quantum state $\rho_{A_1\ldots A_N}$,
the measure of multipartite entanglement, using the FS metric, is defined as 
\begin{equation}
  \mathcal{E}_{\mathcal C}^{FS}(\rho_{A_1\ldots A_N})=\min_{\rho^S_{A_1\ldots A_N} \in {\mathcal S_C} }
  \left( 1 - \frac{\mbox{Tr}\left[\rho_{A_1\ldots A_N} \rho^S_{A_1\ldots A_N} \right]}{\mbox{Tr}\left[\rho_{A_1\ldots A_N}^2\right]} \right)
\label{eqn:mGM}
\end{equation}
where the minimization is carried over a certain class, ${\mathcal S_{\cal C}}$, of separable states.
There is again a hierarchy in the geometric measures defined in this way, just like for pure states.
For example, if the $ \rho^S_{A_1\ldots A_N}$ are fully separable, 
the corresponding measure denoted here as $\mathcal{E}_0^{FS}(\rho_{A_1\ldots A_N})$, can be called ``geometric measure'' of entanglement under the FS metric.
If the minimization is considered over all the  states which are not 
$(k-1)$-separable, then the measure,  denoted as $\mathcal{E}_{N-k+1}^{FS}(\rho_{A_1\ldots A_N})$, 
quantifies the entanglement content of the class of states $\{\rho_{A_1\ldots A_N}\}$ which are $(k-2)$-separable, $(k-3)$-separable and so on upto  
the set of genuinely multipartite entangled states.
Like in the case of pure states, we can also define a valid measure of multipartite entanglement, given by  
  \begin{equation}
  \mathcal{G}\left(\mathcal{E}_\mathcal{C}^{FS} \right)= \cos^{-1} \sqrt{  1-\mathcal{E}_\mathcal{C}^{FS} },
  \label{eqn:gefs}
 \end{equation}
 which is a monotonically increasing function of $\mathcal{E}_\mathcal{C}^{FS}$.
By definition, $\hbar S_0 \geqslant \hbar \mathcal{G}\left(\mathcal{E}_\mathcal{C}^{FS} \right)$. 
Hence 
from Eq. (\ref{eqn:gqurmixfs}),
we obtain
\begin{equation}
 \tau \ \varDelta H_Q \geqslant \hbar \ \mathcal{G}\left(\mathcal{E}_\mathcal{C}^{FS} \right),
\end{equation}
which is an uncertainty relation involving multipartite entanglement and the energy fluctuation of multipartite quantum states involved in a quantum dynamics.

\subsection{Bures metric, GQUR, and entanglement} 

The distance between two arbitrary quantum states can geometrically be measured also by the 
Bures metric, which is defined as
 \begin{equation}
  dS_B^2= \left( 2-2 \sqrt{F \left( \rho(t), \rho(t+dt) \right)}  \right),
 \end{equation}
 where the Uhlmann fidelity, $F$, is given by
 \begin{equation}
  F(\rho_1, \rho_2) = \left( \mbox{Tr} \sqrt{\sqrt{\rho_1} \rho_2 \sqrt{\rho_1}} \right)^2.
 \end{equation}
The metric is Riemannian. Moreover in contrast to the FS metric, it is also contractive under completely positive maps.
In this case, the energy fluctuation, $\varDelta H (t)^2= \mbox{Tr}(\rho H^2) - (\mbox{Tr} \rho H)^2$, can be shown to be 
related to the Uhlmann fidelity \cite{Uhlman92} as
\begin{equation}
 \int_0^{\tau} \varDelta H (t) dt \geqslant \hbar \ \cos^{-1} \sqrt{ F(\rho(0), \rho(\tau)) }.
\end{equation}
where $\rho(0) \ \mbox{and} \ \rho(\tau)$ respectively correspond to the quantum states at times $t=0$ and $t=\tau$. 
For time-independent Hamiltonians, it again reduces to 
\begin{equation}
 \tau \ \varDelta H  \geqslant \hbar \ \cos^{-1} \sqrt{ F(\rho(0), \rho(\tau)) }.
\end{equation}
The above relation is known as the time energy geometric quantum uncertainty relation using Bures metric.

We can exploit the Bures metric to define distance based measure of multipartite entanglement, and 
for an arbitrary N-party quantum state $\rho_{A_1\ldots A_N}$, it reads
\begin{eqnarray}
&&\mathcal{E}_\mathcal{C}^B(\rho_{A_1\ldots A_N})= \nonumber\\
&&\min_{\rho^S_{A_1\ldots A_N}\in \mathcal{S}_C}\left( 1 - \mbox{Tr}\sqrt{\sqrt{\rho^S_{A_1\ldots A_N}}(\rho_{A_1\ldots A_N})  \sqrt{\rho^S_{A_1\ldots A_N}}}  \right).\nonumber\\
\end{eqnarray}
In particular, if the minimum is taken over the set of states which are not genuinely multipartite entangled, the 
corresponding  measure is denoted as $\mathcal{E}_G^B(\rho_{A_1\ldots A_N})$, and quantifies
the genuine multipartite entanglement present in the quantum state.
%
%
%
%
Therefore, time-energy geometric quantum uncertainty relation, in terms of multipartite 
entanglement measure quantified by using the Bures metric, is given by
\begin{equation}
 \tau \ \varDelta H   \geqslant \hbar \ \mathcal{G}\left(\mathcal{E}_\mathcal{C}^{B} \right). 
\end{equation}
A similar relation, but involving the time integral of the energy fluctuation, holds for time-dependent Hamiltonians.

\section{Conclusion}
In quantum theory, the celebrated Heisenberg uncertainty relation provides a bound on the position and momentum uncertainties, in terms of Planck's constant. Indeed, for most of the history of quantum theory, the Planck's constant has played a dominating role in providing the ultimate bounds on our ability to measure two incompatible observables and in a variety of other aspects. 
In the ground-breaking works on Bell inequality and other quantum information tasks, it is quantum entanglement that seems to dominate most of the developments. It is intriguing to ask whether quantum entanglement also sets a fundamental bound on the quantum fluctuations. In this work, for the first time, we have shown that it is not only the Planck's constant but also quantum entanglement that plays an important role in setting the limits for the quantum uncertainties. This is attained by using the geometry of space of quantum states. This underlines the power of geometric ideas in quantum theory 
-- they 
help to bring together two of the most fundamental ingredients of quantum theory, namely, the ``quantum of action'' and the quantum entanglement''. To be specific, we have found a relation between the time-energy uncertainty and the geometric measure of multipartite 
entanglement for both pure and mixed quantum states. The time-energy uncertainty relation is shown to be bounded below by the multipartite entanglement. We find that the minimum time taken for an initial quantum state to reach a target entangled state is directly proportional to the geometric measure of multipartite entanglement.  We have given examples by which our results can be illustrated clearly. We believe that these findings will have important bearing in many areas of quantum theory, including quantum information processing, precession measurements and quantum metrology.

\textit{Acknowledgement}-- R.P. acknowledges support from the Department of Science and Technology, Government of India, in 
the form of an INSPIRE faculty scheme at the Harish-Chandra Research Institute (HRI), India.

\end{document}